\documentclass[prl,twocolumn,superscriptaddress]{revtex4-2}

\usepackage{amsmath,amssymb,amsfonts,bm,color,graphicx,tabularx,physics}
\usepackage{balance}
\usepackage[unicode=true,colorlinks=true]{hyperref}
\usepackage{bibunits}
\defaultbibliography{sandwich-junction}
\usepackage[etex=true,export]{adjustbox}

\hypersetup{linkcolor=blue,citecolor=blue,urlcolor=blue}
\usepackage{diagbox}
\usepackage{color}

\renewcommand{\v}[1]{{\bf #1}}

\newcommand{\beq}{\begin{equation}}
\newcommand{\eeq}{\end{equation}}
\newcommand{\beql}{\begin{equation*}}
\newcommand{\eeql}{\end{equation*}}
\newcommand{\beqn}{\begin{eqnarray}}
\newcommand{\eeqn}{\end{eqnarray}}
\newcommand{\nn}{\nonumber\\}

\renewcommand{\vec}[1]{\mbox{\boldmath$#1$}}

\begin{document}
\bibliographystyle{apsrev4-2}
\title{Meissner effect induced Majorana zero modes at small magnetic field}

\author{Xiao-Hong Pan}
\affiliation{School of Physics and Institute for Quantum Science and Engineering, Huazhong University of Science and Technology, Wuhan, Hubei 430074, China}
\affiliation{Wuhan National High Magnetic Field Center and Hubei Key Laboratory of Gravitation and Quantum Physics, Wuhan, Hubei 430074, China}

\author{Li Chen}
\affiliation{State Key Laboratory of Low Dimensional Quantum Physics, Department of Physics, Tsinghua University, Beijing, 100084, China}

\author{Dong E. Liu}
	\email{dongeliu@mail.tsinghua.edu.cn}
	\affiliation{State Key Laboratory of Low Dimensional Quantum Physics, Department of Physics, Tsinghua University, Beijing, 100084, China}
	
	\author{Fu-Chun Zhang}
	\email{fuchun@ucas.ac.cn}
	\affiliation{Kavli Institute for Theoretical Sciences,  University of Chinese Academy of Sciences, Beijing 100190, China}
	
	\author{Xin Liu}
	\email{phyliuxin@hust.edu.cn}
	\affiliation{School of Physics and Institute for Quantum Science and Engineering, Huazhong University of Science and Technology, Wuhan, Hubei 430074, China}
	\affiliation{Wuhan National High Magnetic Field Center and Hubei Key Laboratory of Gravitation and Quantum Physics, Wuhan, Hubei 430074, China}

\begin{abstract}
 One fundamental difficulty in realizing Majorana zero modes (MZMs) is the required high magnetic field, which causes serious issues, e.g., shrinks the superconducting gap, reduces topological region, and weakens their robustness against disorder. In this work, we propose that the Meissner effect can bring the topological superconducting phase to a superconductor/topological-insulator/superconductor (SC/TI/SC) hybrid system. Remarkably, the required magnetic field strength ($<$10 mT) to support MZMs has been reduced by several orders of magnitude compared to that ($>$0.5 T) in the previous schemes. Tuning the phase difference between the top and bottom superconductors can control the number and position of the MZMs. In addition, we account for the electrostatic potential in the superconductor/topological-insulator (SC/TI) interface through the self-consistent Schrödinger-Poisson calculation, which shows the experimental accessibility of our proposal. Our proposal only needs a small magnetic eld of less than 10 mT and is robust against the chemical potential fluctuation, which makes SC/TI/SC hybrid an ideal Majorana platform.
\end{abstract}
\date{\today}
\maketitle

{\it{Introduction}}-- Unpaired Majorana zero modes (MZMs) can only exist in systems with broken time-reversal symmetry, which usually requires applying a magnetic field. Zeeman splitting is usually used in various strategies~\cite{Sau2010,Lutchyn2010,Oreg2010,Deng2012,Rokhinson2012,Das2012,Wang2012,Churchill2013,Chang2015,Sun2016,Albrecht2016,Wiedenmann2016,Jeon2017,Liu2017,Zhang2017,Zhu2018,Liu2018,Volpez2019,Pan2019,Liu2020,Chen2021,Zhang2021,Papaj2021,Xie2021,Song2022,Li2022,Oshima2022} to break time-reversal symmetry and lead to the topological superconducting phase transition. The Zeeman splitting typically needs to be greater than the superconducting energy gap for this to occur. Therefore a magnetic field above a specific value is the prerequisite in various theoretical proposals and experimental detection. However, a large magnetic field normally suppresses the s-wave superconductivity. At the same time, a sizable and hard superconducting gap is necessary to isolate the MZMs from other low-energy states in the energy space to protect quantum information. As a result, one of the most difficult challenges in implementing MZMs is to balance these two seemingly contradictory prerequisites. A recent experiment~\cite{Zhu2021} observes the Meissner effect induced Doppler shift that can make the superconducting topological surface states have a segmented Fermi surface, a gapless superconducting state, under a small magnetic field about $B_c = 20$~mT. It thus suggests that a small magnetic field can significantly affect superconductivity, prompting us to wonder whether the Meissner effect can induce the topological superconducting phase transition at such a low magnetic field. Besides, a topological superconducting system with controllable number and position of MZMs is a prerequisite for implementing topological qubits. Higher-order topological superconductors~\cite{Langbehn2017,Yan2018,Wang2018,Wang2018a,Hsu2018,Liu2018,Volpez2019,Pan2019,Ezawa2019,Ghorashi2019,Franca2019,Zhang2019,Tiwari2020,Wu2020,Kheirkhah2020,Kheirkhah2022,Wu2022a,Luo2022,Zhu2022,Tan2022} have shown their potential in fulfilling this condition~\cite{Volpez2019,Yan2019,Zhang2020,Zhang2020a,Pahomi2020,Lapa2021,Amundsen2022,Lu2022,Wu2022}.

\begin{figure}[htbp]
    \centering
    \includegraphics[width=1\columnwidth]{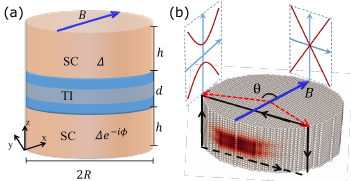}
    \caption{(a) Schematic of the sandwich junction. (b) The density plot of MZMs at $B=0.1 B_c, \phi_{\rm sc}=\pi$. $B_c$ is the critical magnetic field strength to close the top-bottom surface superconducting gap due to the Doppler shift. The two spectrum plots schematically show the $\theta$-dependent edge gap. The black and blue arrows indicate the vector potential under the London gauge and the magnetic field.}
    \label{Figure_1}
\end{figure}
In this work, we demonstrate that Meissner effect can implement the second-order topological superconductivity and controllable Majorana zero modes in the SC/TI/SC sandwich junction under a small magnetic field of less than 10~mT (Fig.~\ref{Figure_1}(a)). An applied magnetic field parallel to the $x-y$ plane induces the opposite diamagnetic current at the top and bottom SC/TI interfaces due to the Meissner effect  (Fig.~\ref{Figure_1}(b)). Remarkably, this diamagnetic current breaks time-reversal symmetry and causes the coexistence of the two different types of mass domain walls at the edge of the SC/TI/SC junction. The mass domain walls have multiple zeros where the unpaired MZMs locate (Fig.~\ref{Figure_1}(b)) even if the applied magnetic field is less than 10mT. Remarkably, the number and positions of MZMs can be controlled by the superconducting phase difference across the junction, the applied magnetic field strength, and the system size. To be concrete, we perform self-consistent Schr\"{o}dinger-Poisson calculation to include the static potential at the SC/TI surface. The Majorana phase transition survives even though the static potential brings the surface Dirac point far away from the Fermi level, indicating our proposal's robustness against the chemical potential fluctuation.

{\it{Continuous model--}} For simplicity, we start with a continuous model to describe the SC/TI/SC junction without the magnetic field as \cite{Liu2011,supp}
\beqn\label{Ham-sys}
H_{0}&=& \hbar v(\hat{p}_{x}s_{y}-\hat{p}_{y}s_{x})\rho_{z}\tau_{z} + (m+t(r))\rho_x \tau_z-\mu \tau_{z}  \nn
&+&\Delta\Big[\frac{(1+\rho_z)}{2} \tau_x+\frac{(1-\rho_z)}{2} (\cos\phi\tau_x+\sin\phi\tau_y)\Big],
\eeqn
with $v$ and $\hat{p}$ the TI surface state velocity
and the momentum operator respectively, $\mu$ the chemical potential, $m$ and $t(\vec{r})$ the hybridization between two layers through bulk and lateral surface, respectively, $\Delta$ the superconducting gap amplitude, $\phi$ the superconducting phase of the bottom superconductor and the Pauli matrices $\rho$, $\tau$, and $s$ acting on the top-bottom surfaces, Nambu and spin space, respectively. This model is valid when the TI thickness $l \ll d< \xi_{\rm TI}$ with $l=\hbar v/M$ and $\xi_{\rm TI} = \hbar v/\Delta$ the TI surface state decay length and coherence length respectively with $M$ the TI bulk gap. For NbSe$_2$/Bi$_2$Se$_3$ hybrid~\cite{Dai2017,Wang2012}, $l \approx 1$ nm and $\xi_{\rm TI} \approx 100 $ nm cause $ m \ll \Delta$. The open boundary condition (OBC) allows to couple the top and bottom TI layers through the lateral surface, resulting in the radius $r$-dependent coupling coefficient $t(r)$ to be finite at the boundary and negligibly small in the bulk \cite{supp}. Here, the system takes circular geometry for convenience. For $\phi=\pi$, SC/TI/SC sandwich possesses helical Majorana edge modes (MEMs) protected by time-reversal symmetry~\cite{Liu2011,supp}. Notably, the intralayer couplings so far respect time-reversal symmetry and will not affect the helical MEMs. 
 When $\phi$ deviates from $\pi$ as $\phi=\pi+\delta \phi$, it is equivalently to add a uniform tunneling phase $\delta \phi/2$ in the interlayer coupling but keep $\phi=\pi$ \cite{supp}. The interlayer coupling now takes \cite{supp}  
\beqn
H_{\rm J} = t(r)\Big[\cos (\frac{\delta\phi}{2})\rho_{x}\tau_{z}-\sin(\frac{\delta\phi}{2})\rho_{y}\tau_{0}\Big],\nonumber
\eeqn
whose projection to the helical edge state subspace becomes
\beqn\label{JE}
V_{\rm J} = -\bar{t} \sin(\frac{\delta\phi}{2}) \tilde{s}_x= \bar{t} \cos(\frac{\phi}{2}) \tilde{s}_x,
\eeqn
with $\tilde{s}$ the Pauli matrix acting on the helical edge state subspace and $\bar{t}$ the expectation value of $t(r)$ on the edge states \cite{supp}. Note that Eq.~\eqref{JE} takes the exact form of the $4\pi$-periodic Josephson effect with $\bar{t}$ the Josephson coupling strength~\cite{Fu2008}.

\begin{figure}[htbp]
    \centering
    \includegraphics[width=1\columnwidth]{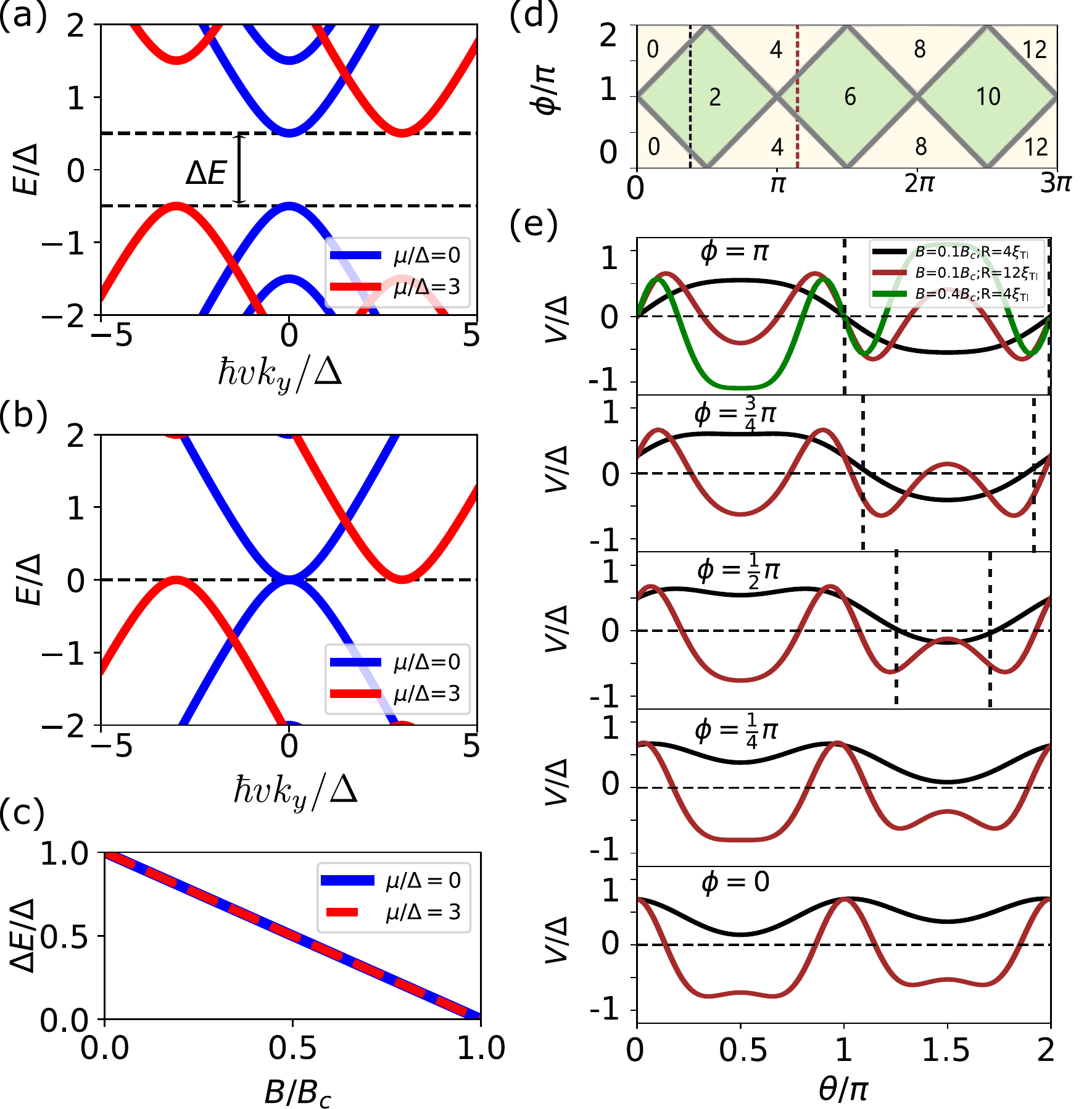}
    \caption{In (a-c), the blue and red curves correspond to $\mu=0$ and $\mu=3\Delta$, respectively. (a) and (b) Energy bands at $B=0.5B_{c}$ and $B=B_{c}$; (c) Band gap as a function of $B$. (d) The number of the band gap zeros for $V_{\rm J}(\theta)$ in the $\eta-\phi$ parameter space. (e) Given $d=\lambda_{L}$ and $\bar{t}=0.7\Delta$, the magnitudes of the mass terms as the function of $\theta$ at $\phi=\pi, 3/4\pi, 1/2\pi,1/4\pi, 0$ from top to bottom.  With varying $B$ and $R$, the black, red, and green curves correspond to $\eta= 1.2, 3.6, 4.8$, respectively. }
    \label{Figure_2}
\end{figure}
Now we apply a magnetic field $B$ along the x-axis. The Meissner effect causes the diamagnetic current parallel and anti-parallel to the y-axis at the top and bottom SC/TI interfaces, respectively (Fig.~\ref{Figure_1}(b)). As the magnetic field considered in this work is small ($< 10$~mT), the superconducting gap is still uniform due to the London rigidity~\cite{London1948,Schrieffer1964}. Therefore, taking the London gauge~\cite{London1935, Bardeen1951}, the diamagnetic current affects the system Hamiltonian only through the vector potential. Firstly, the opposite diamagnetic currents lead to opposite vector potential at the top and bottom SC/TI interfaces (Fig.~\ref{Figure_1}(b)) as $\vec{A}= (0,B\lambda_{L}\rho_z,0)$~\cite{supp} with $\lambda_{L}$ the London penetration depth~\cite{Fletcher2007, Zhu2021}. It modifies the momentum operator $\hat{p}_{y} \rightarrow \hat{p}_y+e A_{y}$ with $e>0$ and gives an additional term \beqn\label{ZM}
H_{\rm A}=-evB\lambda_{L}\rho_0s_x\tau_0,
\eeqn
which functions as an in-plane Zeeman effect and causes the Doppler shift in the x-direction at the top-bottom surfaces as ~\cite{Reinthaler2015,Yuan2018}   
\beqn\label{E_bulk}
E(k_{x}=0)=s_{x}\rho_z evB\lambda_{L}\pm\sqrt{\Delta^{2}+(\mu-s_{x}\rho_z \hbar vk_{y})^{2}}.
\eeqn
The Doppler shift reduces (Fig.~\ref{Figure_2}(a)) and eventually closes the system gap at $B_c = \Delta / (ev\lambda_L)$ (Fig.~\ref{Figure_2}(b)), which remarkably is independent of the chemical potential as shown in Fig.~\ref{Figure_2}(c). Therefore, in a wide range of chemical potential, we can take $B_c = \Delta / (ev\lambda_L)$ as the critical field to close the bulk superconducting gap. The diamagnetic current also affects the vector potential at the lateral surface: When taking OBC with the small magnetic field ($B< B_c$), the flux in the TI region (enclosed by the black rectangular in Fig.~\ref{Figure_1}(b)) can be calculated as 
\beqn
B 2dR\sin\theta=\oint \vec{A} \vec{dl}=2(-B\lambda_{L} 2R\sin \theta + \delta \Phi (\theta)), \nonumber
\eeqn 
where the first and second terms in the above integral come from the vector potential at the top-bottom and lateral surfaces, respectively. Note that the first term contributes to the flux opposite to the total flux in this region (Fig.~\ref{Figure_1}(b)), which gives $\delta \Phi(\theta) = B(2\lambda_L + d)R\sin\theta$ \cite{supp}. This also indicates a larger Doppler shift at the lateral surface and offers an additional $\theta$-dependent phase $-\pi \delta\Phi(\theta)/\Phi_0=-\eta\sin\theta$
 into the electron tunneling from the bottom to the top surfaces through the edge, with $\Phi_0$ the magnetic quantum flux and
\beqn\label{eta}
\eta =  \frac{\pi B R(2\lambda_L+d)}{\Phi_0} = \frac{B}{B_c} \frac{R}{ \xi_{\rm TI}} (2+\frac{d}{\lambda_L})
\eeqn
a unitless parameter, characterizing the ratio between the flux through the SC/TI/SC junction and the magnetic quantum flux. The interlayer coupling through the lateral surface now becomes \cite{supp}
\beqn
H_{\rm J}= t(r)\Big[\cos (\frac{\delta\phi}{2}-\eta\sin\theta)\rho_{x}\tau_{z}-\sin(\frac{\delta\phi}{2}-\eta\sin\theta)\rho_{y}\tau_{0}\Big].\nonumber
\eeqn
Given the polar angle $\theta$, we can project $H_{\rm A}$ of Eq.~\eqref{ZM} into the edge states subspace and obtain the Jackiw-Rebbi Hamiltonian \cite{Jackiw1976, Bernevig2013, Wu2019, supp}
\beqn\label{mass-1}
H_{\rm sf}(\theta)=-i\frac{\hbar v_f}{R} \partial_{\theta}  \tilde{s}_z + \Big(V_{\rm A}(\theta) + V_{\rm J}(\theta) \Big) \tilde{s}_x,
\eeqn 
where 
\beqn\label{mass-2}
V_{\rm A}(\theta) = -\Delta \frac{B}{B_c}\sin\theta, \ \ V_{\rm J}(\theta)= - \bar{t}\sin(\frac{\delta\phi}{2}-\eta\sin\theta)
\eeqn
correspond to $H_{\rm A}$ and $H_{\rm J}$ respectively \cite{supp}. 

\begin{figure}[htbp]
    \centering
    \includegraphics[width=1\columnwidth]{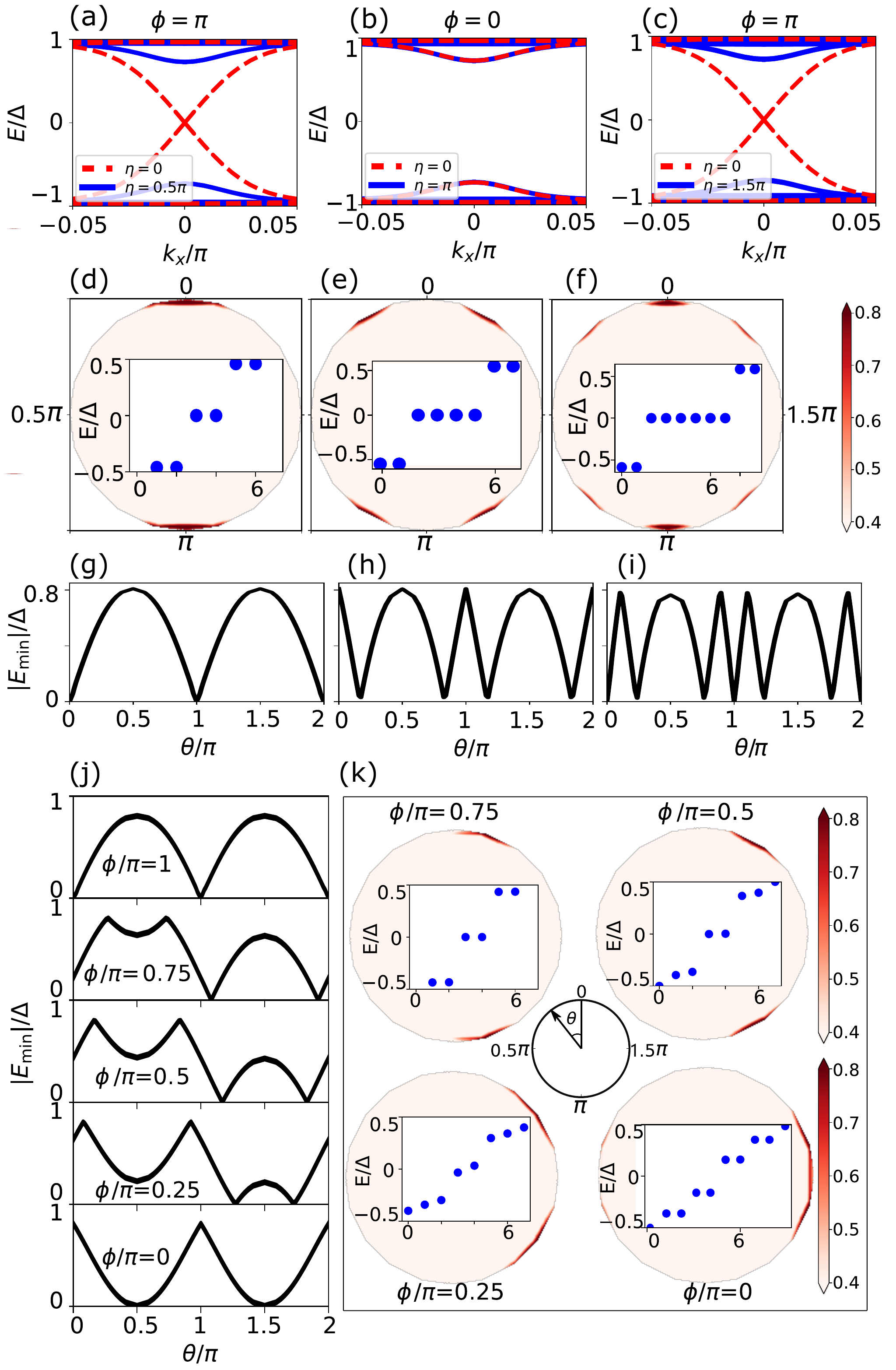}
    \caption{Given $d$ and $R$: (a-c) the energy spectrum for $\theta=\pm \pi/2$ with $\phi=(\pi,0,\pi)$, respectively. The dashed red and solid blue curves correspond to $B=0$ and $B \neq  0$. The latter gives $\eta=(0.5\pi,\pi,1.5\pi)$. (d-e) The MZMs density and several eigenvalues closest to the Fermi level. (g-i) the associated gap functions of the edge states as a function of $\theta$. (j) the lowest positive eigenvalue as a function of $\theta$ for different $\phi$; (k) the MZMs density and several eigenvalues closest to the Fermi level at $\eta=0.5\pi$ for different $\phi$. Parameters: $a=1$ the lattice constant, $d=\lambda_{L}=20$, $R=25$, $t_{x,y,z}=1$, $M=-1.5$, $\alpha_{x,y,z}=2$, $\mu=0$, $\mu_{sc}=1.75$, $t_c=1$; in SC region $t_s=\hbar^{2}/2m_s=1$ for hopping in $z$-direction and $t_s=0$ for hopping in $x$- or $y$-direction; $\Delta=0.1$ in (a-c) and $\Delta=0.3$ in (d-k).}
    \label{Figure_3}
\end{figure}

The $V_{\rm A}$ is equivalent to the Zeeman splitting induced mass term in implementing two high-order Majorana corner states around $\theta=0$ and $\pi$  \cite{Zhu2018,Volpez2019,Zhang2021}. In NbSe$_2$/Bi$_2$Te$_3$ hybrid, the magnetic field of about 10 mT can generate a gap with 0.25 meV, which implies the effective g-factor about 800~\cite{Zhu2021}, much larger than all the g-factor in the current Majorana platform~\cite{Nilsson2009,Stanescu2011,Albrecht2016,Winkler2017,Fornieri2019}. The $V_{\rm J}$ is a nested sine function that can possess multiple zeros at $\delta\phi/2-\eta\sin\theta=n\pi$ with n an integer number and $\theta\in (0,2\pi)$(Fig.~\ref{Figure_2}(d)). For $\eta<\pi$ and $\delta \phi = 0$, it has a similar potential to the first and together with the first term can produce two mass sign changes as indicated by the black curves in Fig.~\ref{Figure_2}(e). When increasing $\eta$ by increasing either the magnetic field (green curve in Fig.~\ref{Figure_2}(e)) or the system size ($R$ or $d$) (brown curve in Fig.~\ref{Figure_2}(e)) according to Eq.~\eqref{mass-1}, there appear more mass sign changes (Fig.~\ref{Figure_2}(e)). Note that each mass sign change indicates the location of the unpaired MZM. When varying $\delta \phi$ and fixing $\eta=1.2$ and $\eta=3.6$, the location and the number of MZMs are changed as shown by the black and brown curve in Fig.~\ref{Figure_2}(e) for $\phi=\pi$, $3\pi/4$, $\pi/2$, $\pi/4$ and $0$. Therefore the second mass term has two advantages over the first one: we can control the MZMs number by varying either the magnetic field, the system size, or the superconducting phase difference; we can achieve unpaired MZMs in an even smaller magnetic field by increasing the system size.

{\it{Lattice model--}} We now proceed to confirm the above analysis in the lattice model. Without applying a magnetic field, the lattice model for the SC/TI/SC sandwich takes the form
\beqn
H=\begin{bmatrix}
    H_{\rm TI}+H_{\rm sc}+H_{\rm c} &  \Delta(z)\\
     \Delta(z)^{*}& -s_{y}(H_{\rm TI}+H_{\rm sc}+H_{\rm c})^{*}s_{y}
\end{bmatrix},
\eeqn
with $\Delta(z)=\Delta$ for $z>d/2$, $\Delta(z)=\Delta e^{-i\phi}$ for $z<-d/2$ and $\Delta(z)=0$ otherwise, $d$ the thickness of TI and $\phi$ the bottom superconductor phase (Fig.~\ref{Figure_1}(a)). The TI Hamiltonian $H_{\rm TI}$ in momentum-space takes the form~\cite{Zhang2009}
\beqn\label{TI-Ham}
H_{\rm TI}=\sum_{i}[M+2t_{i}(1-\cos k_{i})]\sigma_{z}+\alpha_{i}\sin k_{i}\sigma_{x}\tilde{\sigma}_{i}-\mu, \nonumber
\eeqn
with $i=x,y,z$, the Pauli matrices $\sigma, \tilde{\sigma}$ acting on the orbital and intrinsic angular momentum space, $M$, $t_{i}$, $\alpha_{i}$ the bulk gap, kinetic energy, and the spin-orbital coupling strength, respectively. The electronic Hamiltonian of the SC $H_{\rm sc}=\hbar^{2}k^{2}/2m_{s}-\mu_{s}$ with $m_{s}$ and $\mu_{s}$ the effective mass and chemical potential respectively. The coupling between TI and SC at the interface takes $H_{c}=\sum_{z}-t_{c}c^{\dag}_{k,z+1}c_{k,z}+h.c.$ with $t_{c}$ the coupling strength. When applying the magnetic field, the vector potential under London gauge \cite{Tinkham2004} generally takes 
\beqn\label{vector_TI}
A_{y}(y,z)&=&-f(y)g'(z)+2B\lambda_{L}\frac{z}{d},\nn
A_{z}(y,z)&=&f'(y)g(z).
\eeqn
where $g(\pm d/2) =0$ eliminates the $z$ component of the vector potential at the SC/TI interface and $\oint \vec{A}\cdot \vec{dl}=BS_{\rm{r}}$ with the integration along the cross-section edge and $S_{r}$ the cross section area (Fig.~\ref{Figure_1}(b)). To satisfy these two conditions, we takes $f(y)=\bar{B}y^{2}/2,g(z)=\sech^{2}(z/z_0)$ with $d/z_0 \gg 1$ to satisfy the former and $\bar{B}=B(2\lambda_L+d)/2z_0$ to satisfy the latter. 

  In our numerical simulation, we use the package Kwant~\cite{Groth2014}. For simplicity, we fix the TI size ($d$ and $R$) and vary the phase difference $\phi$ and $\eta$ of Eq.~\eqref{eta}. We first take the periodic boundary condition (PBC) in $x$-direction and calculate the edge state spectrum of the cross-section corresponding to $\theta=\pm \pi/2$ (black rectangular in Fig.\ref{Figure_1}(b)). When $\phi=\pi$ and $\phi=0$ without magnetic field ($\eta=0$), we have gapless and gapped edges respectively (red dashed curves in Fig.~\ref{Figure_3}(a-c)). With magnetic field ($\eta\neq 0$), the edge for either $\phi=0$ or $\phi=\pi$ is gapped (blue curves in Fig.~\ref{Figure_3}(a-c)). These are consistent with our continuous model analysis. Note that for all different $\eta$ (equivalently magnetic field), the absolute value of the minimal eigenenergy, $|E_{\rm min}|$, for blue curves in Fig.~\ref{Figure_3}(a-c), shows similar magnitude of about $0.8\Delta$ (Fig.~\ref{Figure_3}(a-c)) which indicates good SC proximity effect in TI. We further show the number of MZMs and their distribution by taking OBC in all three directions in Fig.~\ref{Figure_3}(d-f), corresponding to the system in Fig.~\ref{Figure_3}(a-c). Note that limited by the numerical resources, we remove the two SC layers and add superconducting pairing into the TI region within the $d/3$ thickness from the top and bottom surfaces while keeping other parameters unchanged in MZMs density calculation. The system exhibits two, four, and six MZMs in Fig.~\ref{Figure_3}(d-f), which is consistent with the analytical result in Fig.~\ref{Figure_2}(d). The locations of the MZMs are around the gap zeros as demonstrated by the numerical plot of the gap magnitude as a function of $\theta$ in Fig.~\ref{Figure_3}(g-i). These results confirm that the number of MZMs can be controlled by varying $\eta$ and $\phi$. To show the MZMs move, we further fix $\eta=0.5\pi$ and plot the gap magnitudes in Fig.~\ref{Figure_3}(j) and the MZMs eigenenergies and the distributions in Fig.~\ref{Figure_3}(k) for $\phi=0.75\pi$, $0.5\pi$, $0.25\pi$, $0$, which demonstrate the consistent move of the gap zeros and the MZMs. Our numerical results exhibit the perfect match with our analytical results in Fig.~\ref{Figure_2}(d) and \ref{Figure_2}(e). Remarkably, our results are independent of the specific form of $f(y)$ and $g(z)$\cite{supp}.

\begin{figure}[htbp]
    \centering
    \includegraphics[width=1\columnwidth]{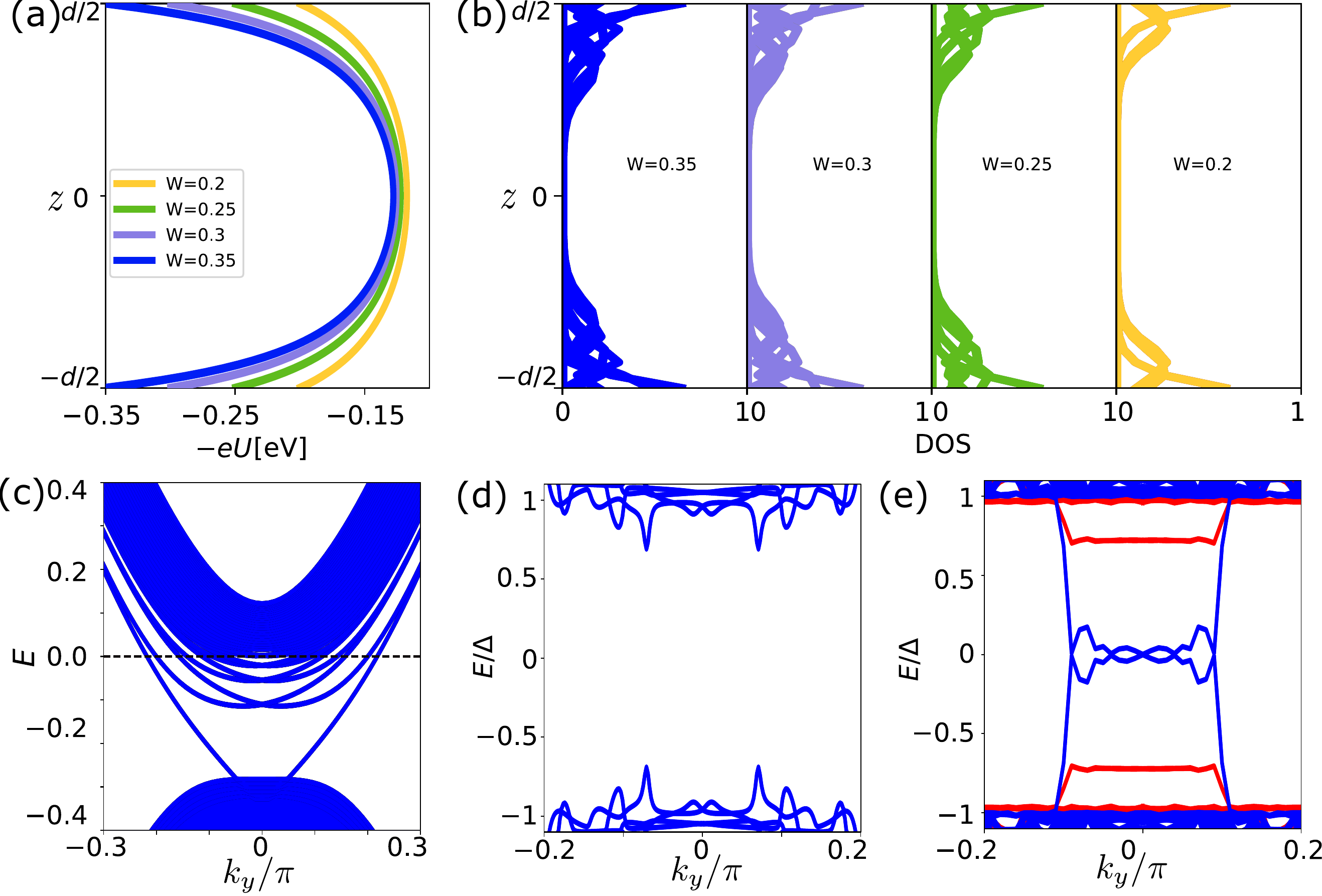}
    \caption{(a) and (b) the static potential $-eU(z)$ and electron density at the Fermi surface with different $W$. (c) and (d) show the electronic and superconducting band dispersion for $(k_x=0,k_y)$ of TI and SC/TI/SC junction, respectively, at $W=0.35$ without a magnetic field. (e) red and blue curves depict the superconducting band dispersion of the cross section for $\theta=\pm \pi/2$ in $y-z$ plane of SC/TI/SC junction with and without a magnetic field respectively. Parameters: $a=1$ nm, $d=30$ nm; $R=1000$ nm; $M=-0.15~{\rm{eV}}; \alpha_{z}=0.22~ {\rm{eV}\cdot nm}; \alpha_{x,y}=0.44~{\rm{eV\cdot nm}}; t_{z}=0.1~{\rm{eV\cdot nm^{2}}}; t_{x,y}=0.566~{\rm{eV}\cdot nm^{2}};\Delta=1.5~ {\rm{meV}}$.}
    \label{Figure_4}
\end{figure}

{\it{Electrostatic potential--}} In the practical scenario, the different work functions between SC and TI lead to the electrostatic field near the SC/TI interface and may deviate the Majorana physics from the ideal models ~\cite{Kiejna1996,Wang2012,Xu2014}. Therefore we calculate the electrostatic potential $U(z)$ in TI using the Schr$\ddot{\rm{o}}$dinger-Poisson method~\cite{Tan1990, Luscombe1992, Ambrosetti2008, Antipov2018}. The parameters of TI Hamiltonian and the relative dielectric constant $\epsilon_{r}=25$ are the values of Bi$_2$Se$_3$~\cite{Zhang2009, Stordeur1992, Chen2022}. We only solve the Schr$\ddot{\rm{o}}$dinger equation in the TI region and treat the top-bottom SCs as the boundary condition $eU(z=\pm d/2)=W$ with $W$ the band offset between the TI and SC~\cite{Mikkelsen2018, Chen2022}. The calculation neglects the magnetic field because it is small. The finite band offset $W$ induces the electrostatic potential near the top and bottom TI surfaces (Fig.~\ref{Figure_4}(a)). We find that the TI spectrum is dramatically affected by the electrostatic potential: at $W=0.35$eV, the Dirac point is embedded deeply into the valence band and more sub-bands other than the surface states appear at the Fermi energy (Fig.~\ref{Figure_4}(b)). Nevertheless, the electrostatic potential confines all the states at the Fermi surface within the 10nm ($\ll \xi_{\rm TI}$) range from the SC/TI interface, indicating good contact with the SCs. To explore the proximity effect on these confined states, we calculate the bulk superconducting spectrum without a magnetic field by taking the PBC in $x$- and $y$-directions and show that all the states at the Fermi surface possess the proximity gap greater than $0.7\Delta$ (Fig.~\ref{Figure_4}(d)). Due to this large proximity gap, at $\phi=\pi$ and taking PBC only in $x$-direction, the SC/TI/SC sandwich shows clean gapless and gapped edge states without and with the magnetic field (Fig.~\ref{Figure_4}(e)). Together, these results indicate that our proposal remains valid in the presence of the electrostatic potential.

{\it{Discussion and Conclusion--}}  In this work, we propose that the Meissner effect induced diamagnetic current can form MZMs in SC/TI/SC sandwich junctions when a small magnetic field ($<10$mT) is present. The diamagnetic current cause the Doppler shifts on TI's top-bottom and lateral surfaces, resulting in two spatially dependent mass terms. The first mass term is equivalent to the Zeeman splitting with a g-factor of up to 800. The second mass term can implement MZMs with controllable numbers and positions. For example, tuning the superconducting phase difference can move the MZMs, providing an experimentally accessible control of MZMs. Further, incorporating the self-consistent Schr$\ddot{\rm{o}}$dinger-Poisson calculation, we show that our results are insensitive to the static potential near the SC/TI interface, which implies no requirement to fine-tune chemical potential. Therefore our proposal has advantages in the material implementation of the Majorana platform.

\begin{acknowledgments}
\section*{Acknowledge}
    We acknowledge useful discussions with Tao Yu, Chao-Xing Liu and Rui-Xing Zhang. X. Liu acknowledges the support of the Innovation Program for Quantum Science and Technology (Grant No. 2021ZD0302700) and the National Natural Science Foundation of China (NSFC) (Grant No.12074133). F.-C. Zhang is partially supported by NSFC grant No. 11674278, and by the Priority Program of Chinese Academy of Sciences, grant No. XDB28000000. D. E. Liu is supported by the Innovation Program for Quantum Science and Technology (Grant No. 2021ZD0302400), the National Natural Science Foundation of China (Grants No. 11974198)
\end{acknowledgments}

\bibliography{sandwich-junction}

\clearpage
\begin{appendix}
\setcounter{figure}{0}
\setcounter{equation}{0}
\renewcommand\thefigure{S\arabic{figure}}
\renewcommand\thetable{S\arabic{table}}
\renewcommand\theequation{S\arabic{equation}}
\begin{widetext}

\begin{center}
 \textbf{Supplement material for ''Meissner effect induced Majorana zero modes at small magnetic field''}
\end{center}

The supplement material contains four sections: A. The inter-layer electronic coupling Hamiltonian through the lateral surface; B. The effective edge Hamiltonian of the BdG system; C. The detail of vector potential distribution; D. Results independent on specific $f(y), g(z)$ function.

\section{A. The inter-layer electronic coupling Hamiltonian through the lateral surface}
For a topological insulator (TI) with finite size, the top and bottom layers of the TI can be coupled through the bulk and lateral surface. The coupling through bulk has been mentioned in a previous study~\cite{Liu2011}. In this section, we mainly study the effective coupling through the lateral surface between the top and bottom layers of the TI.
The TI Hamiltonian $H_{\rm TI}$ in momentum-space, as shown in the main text, takes the form~\cite{Zhang2009}
\beqn\label{TI-Ham}
H_{\rm TI}=\sum_{i}[M+2t_{i}(1-\cos k_{i})]\sigma_{z}+\alpha_{i}\sin k_{i}\sigma_{x} \tilde{\sigma}_{i}-\mu, \nonumber
\eeqn
with $i=x,y,z$, the Pauli matrices $\sigma$ and $\tilde{\sigma}$ acting on the orbital and spin space, $M$, $t_{i}$, $\alpha_{i}$ the bulk gap, kinetic energy, and the spin-orbital coupling strength, respectively. Note that near the Dirac point at each surface, the intrinsic Hilbert space is a two-dimensional subspace expanded by the two states at Dirac point. These two states are also the eigenstate of the associated chiral symmetry operator of the Eq.~\eqref{TI-Ham} for $k_{\parallel}=0$ with $k_{\parallel}$ the in-plane momentum of the surface. Therefore, we can obtain the eigenfunctions
\beqn
\psi_{t}^{1}&=&\ket{\sigma_{y}=+1}\otimes\ket{\tilde{\sigma}_{z}=+1},\nn
\psi_{t}^{2}&=&\ket{\sigma_{y}=-1}\otimes\ket{\tilde{\sigma}_{z}=-1},\nn
\psi_{b}^{1}&=&\ket{\sigma_{y}=-1}\otimes\ket{\tilde{\sigma}_{z}=+1},\nn
\psi_{b}^{2}&=&\ket{\sigma_{y}=+1}\otimes\ket{\tilde{\sigma}_{z}=-1},
\eeqn 
for the top and bottom surfaces and 
\beqn
\psi_{r}^{1}&=&\ket{\sigma_{y}=+1}\otimes\ket{\tilde{\sigma}_{x}=1}+i\ket{\sigma_{y}=-1}\otimes\ket{\tilde{\sigma}_{x}=-1},\\
\psi_{r}^{2}&=&\ket{\sigma_{y}=+1}\otimes\ket{\tilde{\sigma}_{x}=+1}-i\ket{\sigma_{y}=-1}\otimes\ket{\tilde{\sigma}_{x}=-1},
\eeqn
for the lateral surface.

\begin{figure}[htbp]
    \centering
\includegraphics[width=0.8\columnwidth]{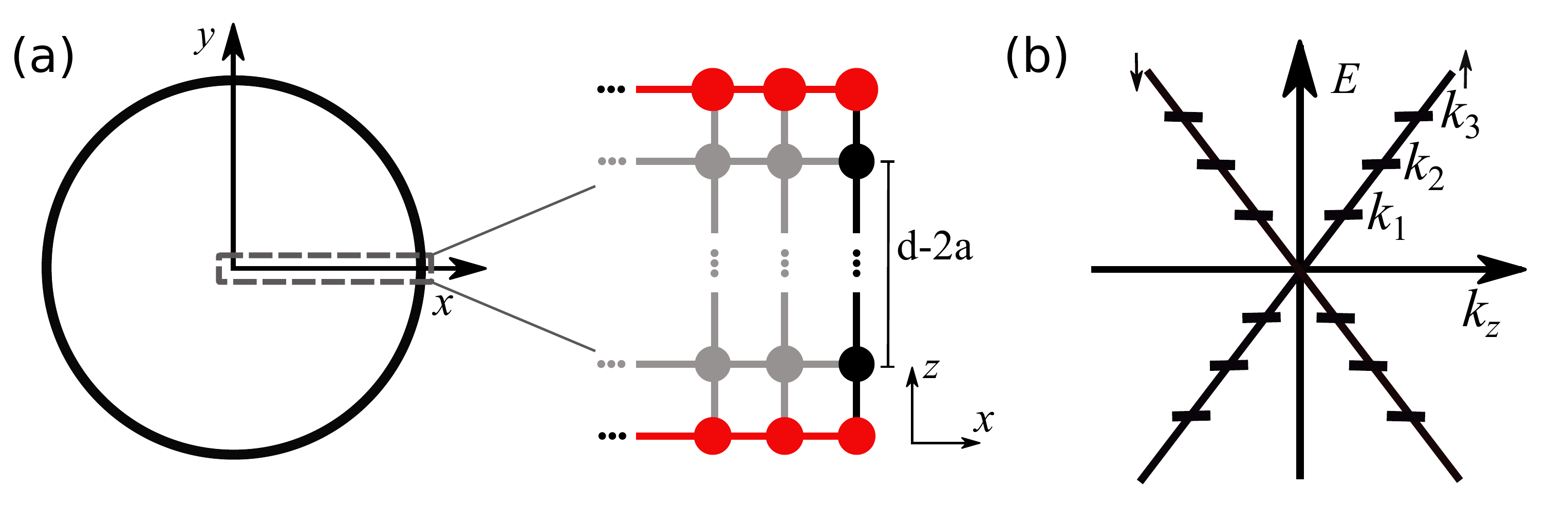}
    \caption{(a) Schematic of the topological insulator. The red and black circles in the x-z plane represent the sites at the top-bottom surface and $x=R$ lateral surface respectively. The gray circles represent the sites of the bulk. (b) Schematic of the surface energy spectrum $E=\hbar v k_{z}$ on the $x=R$ side.  }
    \label{SFigure_3}
\end{figure}

For simplicity, we assume the top and bottom surface states are localized only at the top and bottom layers (red sites in Fig.~\ref{SFigure_3}). They are directly coupled to the lateral surface through the hopping in $z$-direction.
Without loss generality, we first study the effective inter-layer coupling through the lateral surface $x=R$ with the tangent momentum $k_{y}=0$ (Fig.~\ref{SFigure_3}(a)). In this case,
the low energy direct coupling Hamiltonian between the lateral surface and top-bottom surface can be expressed as
\beqn
H_{cp}=\sum_{i,j}T_{t}^{i,j}~c_{t,i}^{\dag}c_{r,j}+T_{d}^{i,j}~c_{r,i}^{\dag}c_{b,j}+h.c.
\eeqn
with
\beqn
&&T_{t}=\sum_{i,j=1}^{2}\bra{\psi_{t}^{i}}-t_{z}\sigma_{z}s_{0}-i\frac{\alpha_{z}}{2}\sigma_{x}s_{z}\ket{\psi_{r}^{j}}=-\frac{1}{2}(t_{z}+\alpha_{z})\begin{bmatrix}
    i &-i \\
    1 & 1 
\end{bmatrix}; \\
&&T_{b}=\sum_{i,j=1}^{2}\bra{\psi_{r}^{i}}-t_{z}\sigma_{z}s_{0}-i\frac{\alpha_{z}}{2}\sigma_{x}s_{z}\ket{\psi_{b}^{j}}=-\frac{1}{2}(t_{z}+\alpha_{z})\begin{bmatrix}
    1&i \\
    1 &-i
\end{bmatrix}.
\eeqn 
and the lateral surface Hamiltonian takes
\beqn\label{Ham_sur_z}
\tilde{H}_{z}=\hbar vk_{z}\begin{bmatrix}
    1 &0 \\
    0 &-1
\end{bmatrix}
\eeqn
in the basis $(\psi_{r}^1,\psi_{r}^2)^{\rm T}$. Therefore, the effect of the lateral surface on the top and bottom system can be accounted in the self-energy
\beqn
\Sigma=\begin{bmatrix}
    0 &\Sigma_{tb} \\
    \Sigma_{bt} & 0 
\end{bmatrix},
\eeqn
where
\beqn\label{Ham-T}
\Sigma_{tb}=T_{t} \big(\sum_{n}\frac{\vert \psi_{n}^1(z=d/2-a)\rangle \langle \psi_{n}^1(z=-d/2+a)\vert + \vert \psi_{n}^2(z=d/2-a)\rangle \langle \psi_{n}^2(z=-d/2+a)\vert}{-\epsilon_{k_{z},n}+i\delta}\big) T_{b}
\eeqn
with $a$ the lattice constant, $k_z$ is quantized as $k_z \in (2n\pi/(d-a))$ taking the periodic boundary condition and $\epsilon_{k_{z},n}$ and $\psi^{1(2)}_{n}$ the eigenvalue and eigenfunctions in Eq.~\eqref{Ham_sur_z}. Note that the Hermitian part of the self-energy affects the Hamiltonian and the anti-Hermitian part modifies the spectral function. Therefore, the effective coupling Hamiltonian between the top and bottom surface can be obtained via accounting for the Hermitian part of the self-energy and takes the form 
\beqn
H_{cp} = \begin{bmatrix}
    0 &H_{tb} \\
    H_{tb}^{\dagger} & 0 
\end{bmatrix}
\eeqn
with 
\beqn
H_{tb} &=& -T_{t} \Bigg(\sum_{n \neq 0}\frac{\vert \psi_{n}^1(z=d/2-a)\rangle \langle \psi_{n}^1(z=-d/2+a)\vert + \vert \psi_{n}^2(z=d/2-a)\rangle \langle \psi_{n}^2(z=-d/2+a)\vert}{-\epsilon_{k_{z},n}}\Bigg) T_{b} \nonumber \\
&=& -T_t \Bigg(\sum_{n>0}\frac{1}{-\epsilon_{k_{n}}}\begin{bmatrix}
    e^{ik_{n}(d-2a)} & 0\\
    0 & e^{-ik_{n}(d-2a)} 
\end{bmatrix} + \frac{1}{\epsilon_{k_{n}}}\begin{bmatrix}
    e^{-ik_{n}(d-2a)} & 0\\
    0 & e^{ik_{n}(d-2a)} 
\end{bmatrix}\Bigg) T_b\nn
&=&-T_t\Bigg(\sum_{n>0}\frac{2i\sin {k_{n}(d-2a)}}{\epsilon_{k_{n}}}\begin{bmatrix}
    -1 & 0\\
    0 & 1 
\end{bmatrix} \Bigg) T_b = t\begin{bmatrix}
    1 & 0 \\
    0 & 1
\end{bmatrix}
\eeqn
with $t=(t_{z}+\alpha)^{2}\sum_{n}\sin {k_{n}a}/\epsilon_{k_{n}}$ ($k_{n}=2n\pi/(d-a)$). Therefore, 
\beqn
H_{cp} = t \begin{bmatrix}
    0 &H_{tb} \\
    H_{tb}^{\dagger} & 0 
\end{bmatrix} =  t \begin{bmatrix}
    0 & 0 & 1 & 0 \\
    0 & 0 & 0 & 1 \\
    1 & 0 & 0 & 0 \\
    0 & 1 & 0 & 0
\end{bmatrix}.
\eeqn
As the system respects rotational symmetry along the z-direction, thus the effective coupling between the top and bottom layers at arbitrary direction $\theta$ is the same as at $x=R$. Therefore, generally speaking, the coupling strength $t$ should only be a function $r$.

\section{B. The effective edge Hamiltonian of the BdG system}\label{Appendix A}
In this section, we provide more details on the calculation of gapless edge states and the corresponding effective edge states Hamiltonian. In BdG basis 
\beqn\label{BdG_basis}
(c_{t,\uparrow},c_{t,\downarrow},c_{b,\uparrow},c_{b,\downarrow},-c_{t,\downarrow}^{\dag},c_{t,\uparrow}^{\dag},-c_{b,\downarrow}^{\dag},c_{b,\uparrow}^{\dag}),
\eeqn
the low energy Hamiltonian of the superconductor/topological insulator/superconductor (SC/TI/SC) junction with superconducting pairing phase $\phi$ between the two SCs takes
\beqn\label{JJ_Ham}
H_{0}&=&v(\hat{p}_{x}s_{y}-\hat{p}_{y}s_{x})\rho_{z}\tau_{z} + (m+t_{0}(\hat{p}_x^2+\hat{p}_y^2)+t(r))\rho_x s_0 \tau_z  -\mu\tau_{z} + \Delta\Big[\frac{(1+\rho_z)}{2} \tau_x+\frac{(1-\rho_z)}{2} (\cos\phi\tau_x+\sin\phi\tau_y)\Big] \nn
\eeqn
with the Pauli matrices $\rho$, $\tau$ and $s$ acting on the top-bottom surfaces, Nambu and spin space, respectively, $v$ the surface state velocity, $\hat{p}_{x(y)}$ the momentum operator, $\mu$ the chemical potential, $\Delta$ the superconducting gap size, $m, t_0$ indicating the tunneling strength between two layers through bulk. Considering the open boundary condition (OBC), the top and bottom TI layers can be coupled through the lateral surface, resulting in the radius r-dependent finite coupling coefficient t(r) at the boundary and negligibly small in the bulk.

According to a previous study~\cite{Liu2011}, the junction possesses gapless helical edge states protected by time-reversal symmetry at $\phi=\pi$. To study the effect of $\phi$ deviating from $\pi$ and magnetic field on the edge state, we first calculate the wave function of the edge state. To simplify the process of calculation, we focus on $\mu=0$ and regard the lateral surface coupling term as the perturbation. The Hamiltonian at $\phi=\pi$ without lateral surface coupling can be simplified as
\beqn\label{Ham-sys}
H_{0}=\tau_{z}\rho_{z}(\hat{p}_{x}s_{y}-\hat{p}_{y}s_{x}) + (m+t_{0}(\hat{p}_x^2+\hat{p}_y^2))\tau_z \rho_x s_0 -\mu\tau_{z}+\Delta \tau_x \rho_z .
\eeqn
It is easy to obtain that Eq.~\eqref{Ham-sys} commutes with operator $\hat{O}_{\rm B}=\tau_{y}\rho_y$ so that we can obtain the block diagonalized Hamiltonian by projecting Eq.~\eqref{Ham-sys} into the eigenstates of $\hat{O}_{\rm B}$. We define a unitary transformation
\beqn
U=(\psi_{\uparrow}^{+,1},\psi_{\downarrow}^{+,1},\psi_{\uparrow}^{+,2},\psi_{\downarrow}^{+,2},\psi_{\uparrow}^{-,1},\psi_{\downarrow}^{-,1},\psi_{\uparrow}^{-,2},\psi_{\downarrow}^{-,2})
\eeqn
with $\psi_{s}^{\pm,i}$ the $i$-th eigenstate of $\hat{O}_{\rm B}$ correspond to eigenvalue $\pm 1$ which takes the form as
\beqn
\psi_{\uparrow(\downarrow)}^{+,1}&=&\frac{1}{\sqrt{2}}\begin{bmatrix}
  0\\1\\1\\0  
\end{bmatrix}\otimes\ket{s_{z}=+(-)}; ~~~~~\psi_{\uparrow(\downarrow)}^{+,2}=\frac{1}{\sqrt{2}}\begin{bmatrix}
  -1\\0\\0\\1  
\end{bmatrix}\otimes\ket{s_{z}=+(-)};\nn
\psi_{\uparrow(\downarrow)}^{-,1}&=&\frac{1}{\sqrt{2}}\begin{bmatrix}
  -1\\0\\0\\-1  
\end{bmatrix}\otimes\ket{s_{z}=+(-)}; ~~~\psi_{\uparrow(\downarrow)}^{-,2}=\frac{1}{\sqrt{2}}\begin{bmatrix}
  0\\1\\-1\\0  
\end{bmatrix}\otimes\ket{s_{z}=+(-)};
\eeqn
The Hamiltonian after transformation gives
\beqn
H'_{0}=U^{\dag}H_{0}U=\begin{bmatrix}
    H_{0}^{+} & 0\\ 0 & H_{0}^{-}
\end{bmatrix}
\eeqn
where
\beqn
H_{0}^{+} &=& -(\Delta+m+t_0(\hat{p}_x^2+\hat{p}_y^2))\tilde{\tau}_{x}-vp_{x}\tilde{\tau_{z}}s_y+vp_{y}\tilde{\tau_{z}}s_{x};\\
H_{0}^{-} &=& (\Delta-m-t_{0}(\hat{p}_x^2+\hat{p}_y^2))\tilde{\tau}_{x}+vp_{x}\tilde{\tau}_{z}s_y-vp_{y}\tilde{\tau}_{z}s_{x};
\eeqn
with Pauli matrix $\tilde{\tau}, \tilde{s}$ acting in effective layer and spin space repestively. After Fourier transformation, $H_{0}^{\pm}$ can be expressed as 
\beqn\label{Lat_Ham}
H_{0}^{+}(k)&=&-(\Delta+m+2t_{0}(2-\cos k_x-\cos k_y))\tilde{\tau}_{x}-v\sin{k_{x}}\tilde{\tau}_z s_y +v \sin{k_{y}}\tilde{\tau}_{z}s_x\nn
H_{0}^{-}(k)&=&(\Delta-m-2t_{0}(2-\cos k_x-\cos k_y))\tilde{\tau}_{x}+v\sin{k_{x}}\tilde{\tau}_z s_y -v \sin{k_{y}}\tilde{\tau}_{z}s_x 
\eeqn
Both of $H_{0}^{+}$ and $H_{0}^{-}$ possess inversion ($\mathcal{I}=\tilde{\tau}_{x}$), time-reversal symmetry ($\mathcal{T}=-is_{y}\mathcal{K}$), and particle-hole symmetry ($\mathcal{P}=\tilde{\tau}_{y}s_{y}\mathcal{K}$). Considering $m=0$ and $\Delta>0$, according to Fu-Kane criteria~\cite{Fu2007}, in the region $t_{0}<-\Delta/4$ ($t_{0}>\Delta/4$), $H_{0}^{+}$ and $H_{0}^{-}$ are topological nontrivial (trivial) and trivial (nontrivial), respectively. Thus, considering $t_{0}>\Delta/4$, we can obtain eigen-states with zero energy in $H_{0}^{-}$ with mixed boundary condition as numerical plot in Fig.~\ref{SFigure_1}. In the following, we first focus on the edge parallel to the x-direction and then extend to an arbitrary direction.
\begin{figure}[htbp]
    \centering
\includegraphics[width=1\columnwidth]{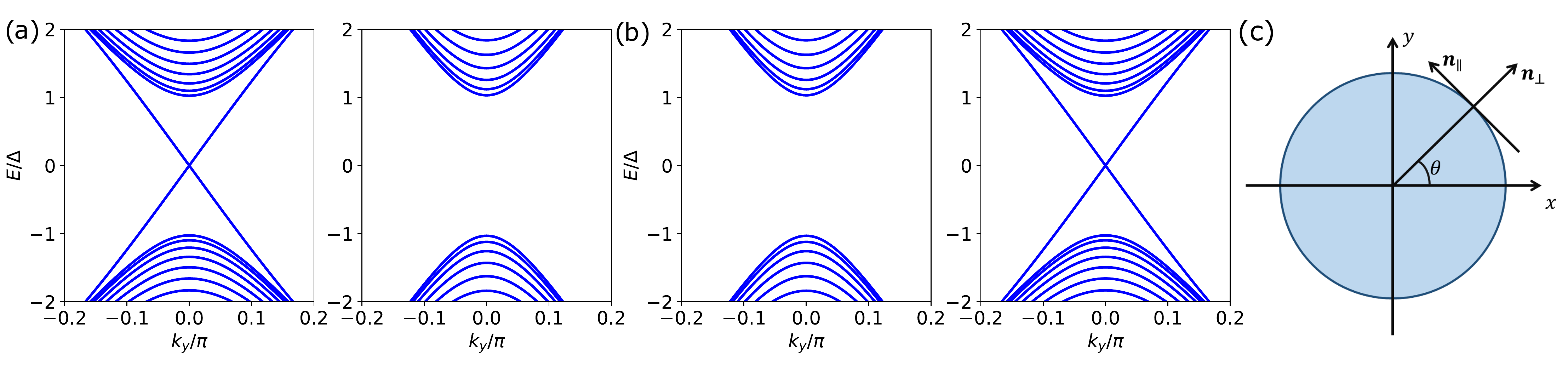}
    \caption{Considering (a) $t_{0}=-1, \Delta=0.5$ and (b) $t_{0}=1, \Delta=0.5$, the left and right figures in (a-b) are the energy spectrum of $H_{0}^{+}$ and $H_{0}^{-}$ with open (periodic) boundary condition along x (y), respectively. (c) Schematic figure of a vector normal and parallel to the edge at $\theta$.}
    \label{SFigure_1}
\end{figure}

In the following, we mainly 
 consider the helical edge states only generated in $H_{0}^{+}$. To obtain the wave function of the edge states easier, we expand the lattice
Hamiltonian (Eq.~\eqref{Lat_Ham}) to second order around $\vec{k}=(0,0)$ which can be expressed as
\beqn
H_{0}^{+}=-(\Delta+m+t_{0}(k_{x}^{2}+k_{y}^{2}))\tilde{\tau}_{x}-vk_{x}\tilde{\tau}_{z}s_{y}+vk_{y}\tilde{\tau}_{z}s_{x}
\eeqn
We first consider the semi-infinite region $x<0$ corresponding to polar angle $\theta=0$ in the main text. We can replace $k_x\rightarrow-i\partial_{x}$ and regard $k_{y}$ as perturbation. The Hamiltonian realizes zero modes at $k_{y}=0$ can be formed as
\beqn\label{Hedge_x}
\tilde{H}_{0}^{+}(k_x\rightarrow-i\partial_{x},k_{y}=0)=-(\Delta+m-t_0\partial_{x}^{2})\tilde{\tau}_x+iv\partial_x\tilde{\tau}_{z}s_{y}.
\eeqn
Obviously, Eq.~\eqref{Hedge_x} anti-commute with chiral symmetry $\mathcal{C}_{x}=\tilde{\tau}_{y}s_{y}$. Thus, the wave function of the edge states with zero energy is also the eigenstates of  $\mathcal{C}_{x}$. In the semi-infinity region, the wave function of the zero modes satisfying boundary condition $\Psi_{i}(0)=\Psi_{i}(-\infty)=0 (i=1,2)$ given by~\cite{Liu2011, Yan2018}
\beqn
\psi_i^{x}=\frac{1}{N_{0}}\sin{\kappa_{1}x}e^{-\kappa_{2}x}\chi_{i}.
\eeqn
with $N_{0}$ normalization value,  $\kappa_{1(2)}$ describe the distribution of the wave-function, $\chi_{i}$ the eigenstate of $\mathcal{C}_{x}$ satisfying $\mathcal{C}_{x}\chi_{i}=-\chi_{i}$ formed as
\beqn
\chi_{1}&=&\ket{\tilde{\tau}_{y}=1}\otimes\ket{s_{y}=-1}=\begin{bmatrix}
    1 \\i
\end{bmatrix}\otimes\begin{bmatrix}
    1\\ -i
\end{bmatrix}; \nn
\chi_{2}&=&\ket{\tilde{\tau}_{y}=-1}\otimes\ket{s_{y}=1}=\begin{bmatrix}
    1 \\-i
\end{bmatrix}\otimes\begin{bmatrix}
    1\\i
\end{bmatrix}.
\eeqn
Substitute the wave-function into Eq.~\eqref{Hedge_x}, we can obtain
\beqn
\kappa_{1}=\sqrt{|(\Delta+m)/t_{0}|-v^{2}/4t_{0}^{2} }; \ \ \kappa_{2}=-\frac{v}{2t_{0}} \ \ \text{and} \ \ |N_{0}|^{2}=4|\kappa_{2}|(\kappa_{1}^{2}+\kappa_{2}^{2})/\kappa_{1}^{2}.
\eeqn
To obtain the wave-function of the edge states $\psi_{1(2)}$ at the arbitrary direction, we define the vector that is normal and parallel to the edge
\beqn
\vec{n}_{\perp}=\cos\theta \vec{x}+\sin\theta \vec{y}; \ 
 \ \vec{n}_{\parallel}=-\sin\theta \vec{x}+\cos\theta \vec{y}
\eeqn
with $\theta$ the angle between the edge and the x-axis as shown in Fig.~\ref{SFigure_1}(c). Now the Hamiltonian in the semi-infinity region can be expressed as
\beqn\label{Hedge_theta}
\tilde{H}_{0}^{+}(k_{\perp}\rightarrow-i\partial_{\perp},k_{\parallel=0})=-(\Delta+m-t_0\partial_{\perp}^{2})\tilde{\tau}_x+iv\partial_{\perp}\tilde{\tau}_{z}s_{y}.
\eeqn
with $s_{\perp}=-\sin\theta s_{x}+\cos\theta s_{y}$ and $s_{\parallel}=\cos\theta s_{x}+\sin\theta s_{y}$. Apparently, Eq.~\eqref{Hedge_theta} possess chiral symmetry $\mathcal{C}_{\theta}=\tilde{\tau}_{y}s_{\perp}$. The edge states wave-function can be obtained by solving the eigenvalue equation
\beqn
\bar{H}_{0}^{+}(k_{\perp}\rightarrow-i\partial_{\perp},k_{\parallel=0})\psi_{1(2)}=0,
\eeqn
where the wave-functions $\psi_{1(2)}$ are also the eigen-state of $\mathcal{C}_{\theta}$ satifing boundary condition $\psi_{+(-)}(0)=\psi_{+(-)}(-\infty)=0$. The wave function of the edge state now takes the form as
\beqn
\chi_{1}&=&\ket{\tilde{\tau}_{y}=1}\otimes\ket{s_{\perp}=-1}=\frac{1}{2}\begin{bmatrix}
    1 \\i
\end{bmatrix}\otimes\begin{bmatrix}
    e^{-i\frac{\theta}{2}} \\-ie^{i\frac{\theta}{2}}
\end{bmatrix};\nn
\chi_{2}&=&\ket{\tilde{\tau}_{y}=-1}\otimes\ket{s_{\perp}=+1}=\frac{1}{2}\begin{bmatrix}
    1 \\-i
\end{bmatrix}\otimes\begin{bmatrix}
    e^{-i\frac{\theta}{2}} \\ie^{i\frac{\theta}{2}}
\end{bmatrix}.
\eeqn
with $\chi_{1(2)}$ the eigenstate of $\mathcal{C}_{\theta}$ satisfying $\mathcal{C}_{\parallel}\chi_{1(2)}=-\chi_{1(2)}$.
We define 
\beqn
\tilde{\psi}_{1}^{\theta}=\frac{1}{\sqrt{2}}(\psi_{1}^{\theta}-i\psi_{2}^{\theta}) \ \ \text{and} \ \  \tilde{\psi}_{2}^{\theta}=\frac{1}{\sqrt{2}}(\psi_{1}^{\theta}+i\psi_{2}^{\theta}) 
\eeqn
which can be related to each by time-reversal symmetry $\mathcal{T}\tilde{\psi}_{1(2)}^{\theta}=\mp\tilde{\psi}_{2(1)}^{\theta}$ and satisfy particle-hole symmetry $\mathcal{P}\tilde{\psi}_{1(2)}^{\theta}=\mp i\tilde{\psi}_{1(2)}^{\theta}$. Projecting $k_{\parallel}\neq0$ term into the edge states $(\tilde{\psi}_{1}^{\theta},\tilde{\psi}_{2}^{\theta})^{\mathcal{T}}$, we can obtain the linear effective edge Hamiltonian
\beqn
H_{\rm sf}=vk_{\parallel}\tilde{s}_{z}.
\eeqn
Since the lateral surface coupling term
\beqn
U^{\dag}t(r)\tau_{z}\rho_{x}U=t(r)\begin{bmatrix}
    -1 & 0\\
    0 & -1
\end{bmatrix}\otimes\tilde{\tau}_{x}s_0
\eeqn
respects the time-reversal symmetry and anti-commutes with chiral symmetry, the projection into the edge state Hilbert space is zero. For $\phi$ deviates from $\pi$ as $\phi=\pi+\delta$, the deviation term breaks time-reversal symmetry and results in non-zero projection. Expanding $\phi$ around $\pi$ to the first order, the derivation term takes the form of $-\Delta\delta\phi\tau_{y}(1-\rho_{z})/2$ and the corresponding projection into the eigenstate of $\hat{O}_{B}$ can be expressed as
\beqn
H'_{\Delta}=U^{\dag}\Big[-\Delta\delta\phi\tau_{y}\frac{1-\rho_{z}}{2}\Big]U=-\Delta\frac{\delta\phi}{2}\begin{bmatrix}
    \tilde{\tau}_{y}s_{0} & i\tilde{\tau}_{z}s_{0} \\
    -i\tilde{\tau}_{0}s_{0} &\tilde{\tau}_{y}s_{0}
\end{bmatrix}.
\eeqn
Projecting $H'_{\Delta}$ into the edge states space, we can obtain that
\beqn
\sum_{i,j}\bra{\tilde{\psi}_{i}^{\theta}}-\Delta\frac{\delta\phi}{2}\tilde{\tau}_{y}s_{0}\ket{\tilde{\psi}_{j}^{\theta}}=-\Delta\frac{\delta\phi}{2}\tilde{s}_{x}\approx-\Delta\sin\Big(\frac{\delta\phi}{2}\Big)\tilde{s}_{x}.
\eeqn
Equivalently, we can add a uniform tunneling phase $\delta\phi/2$ into the interlayer
coupling but keep $\phi=\pi$. Now, the interlayer coupling term takes the form as
\beqn
H'_{\rm J} = U^{\dag}t(r)\Bigg[\cos (\frac{\delta\phi}{2})\tau_{z}\rho_{x}-\sin(\frac{\delta\phi}{2})\tau_{0}\rho_{y}\Big] U=t(r)\Bigg[-\cos (\frac{\delta\phi}{2})\begin{bmatrix}
    \tilde{\tau}_{x}s_{0} & 0\\
    0 & \tilde{\tau}_{x}s_{0}
\end{bmatrix}-\sin(\frac{\delta\phi}{2})\begin{bmatrix}
  \tilde{\tau}_{y}s_{0} & 0\\
    0 & -\tilde{\tau}_{y}s_{0}  
\end{bmatrix}\Bigg]
\eeqn
and the corresponding projection which can be simplified as
\beqn
\sum_{i,j}\bra{\tilde{\psi}_{i}^{\theta}}-\sin(\frac{\delta\phi}{2})\tilde{\tau}_{y}s_{0} \ket{\tilde{\psi}_{j}^{\theta}}=-\sin(\frac{\delta\phi}{2})\frac{1}{N_{0}^{2}}\int\sin^{2}{\kappa_{1}r}e^{-2\kappa_{2}r}t(r)\tilde{s}_{x}=-\bar{t}\sin(\frac{\delta\phi}{2})\tilde{s}_{x}
\eeqn
Now, we consider the magnetic field along the x direction. The magnetic field induces the opposite
diamagnetic currents at the SC/TI interface that lead to opposite vector potential formed as $\vec{A}=(0, B\lambda_{L}\rho_{z},0)$ with $\lambda_{L}$ the penetration length. The vector potential modifies the momentum operator as $\hat{p}_{y}=\hat{p}_{y}+eA_{y}$ with $e>0$ and thus giving additional term
\beqn
H'_{A}=U^{\dag}(-evB\lambda_{L}\tau_{0}\rho_{0}s_{x})U=-evB\lambda_{L}\begin{bmatrix}
    \tilde{\tau}_{0}s_{x} & 0\\
    0  & \tilde{\tau}_{0}s_{x}
\end{bmatrix}.
\eeqn
Projecting $H_{A}$ into edge states space, we can obtain the correspondent mass term
\beqn
\sum_{i,j}\bra{\tilde{\psi}_{i}^{\theta}}-evB\lambda_{L}\tilde{\tau}_{0}s_{x}\ket{\tilde{\psi}_{j}^{\theta}}=-evB\lambda_{L}\sin\theta s_{x}
\eeqn
Besides, the opposite diamagnetic current at the SC/TI interface also affects the vector potential distribution at the lateral surface. According to the integral of the closed black rectangular in the main text, the lateral surface tunneling from the bottom to the top layer obtains additional $\theta$-dependent phase $-\eta\sin\theta$ with $ \eta=\pi BR(2\lambda_{L}+d)/\Phi_{0}$. The block of interlayer coupling through the lateral surface with the eigenvalue -1 of $\hat{O}_{B}$ now changes to 
\beqn
H'_{J}=t(r)\Big[-\cos (\frac{\delta\phi}{2}-\eta\sin\theta)\begin{bmatrix}
    \tilde{\tau}_{x}s_{0} & 0\\
    0 & \tilde{\tau}_{x}s_{0}
\end{bmatrix}-\sin(\frac{\delta\phi}{2}-\eta\sin\theta)\begin{bmatrix}
  \tilde{\tau}_{y}s_{0} & 0\\
    0 & -\tilde{\tau}_{y}s_{0}  
\end{bmatrix}
\Big]
\eeqn
and the projection change to
\beqn
\sum_{i,j}\bra{\tilde{\psi}_{i}^{\theta}}t(r)\Big[-\cos (\frac{\delta\phi}{2}-\eta\sin\theta)
    \tilde{\tau}_{x}s_{0}-\sin(\frac{\delta\phi}{2}-\eta\sin\theta)\tilde{\tau}_{y}s_{0}\Big]\ket{\tilde{\psi}_{j}^{\theta}}=-\bar{t}\sin(\frac{\delta\phi}{2}-\eta\sin\theta) \tilde{s}_{x}
\eeqn
To sum up, the above calculation, considering $\phi$ deviate from $\pi$ and adding an in-plane magnetic field, the effective Hamiltonian of the edge takes the form of Jackiw-Rebbi Hamiltonian~\cite{Jackiw1976, Bernevig2013, Wu2019}
\beqn\label{Ham_edge}
H_{\rm sf}=vk_{\parallel}\tilde{s}_{z}+(V_{A}+V_{J})\tilde{s}_{x}
\eeqn
with 
\beqn
V_{A}=-evB\lambda_{L}\sin\theta \  \ \text{and}   \  \  V_{J}=-\bar{t}\sin(\frac{\delta\phi}{2}-\eta\sin\theta) \nonumber 
\eeqn
which shows the same form as shown in the main text. $H'_{\rm sf}$ is a Dirac equation with a mass sign change at critical direction $\theta_{c}$. We expect to obtain Majorana zero modes at the mass sign change and the ansatz solution of the zero modes take the form as
\beqn
\bar{\Psi}(\theta)=\exp(-\frac{R}{v}\int_{\theta_{c}}^{\theta}V_{A}(\theta')+V_{J}(\theta')d\theta')\bar{\chi}.
\eeqn
Replacing $k_{\parallel}\rightarrow-i\frac{1}{R}\partial_{\theta}$ and acting into Eq.~\eqref{Ham_edge}, we can obtain that
\beqn
\begin{bmatrix}
i(V_{A}(\theta)+V_{J}(\theta)) & (V_{A}(\theta)+V_{J}(\theta)) \\
(V_{A}(\theta)+V_{J}(\theta)) &-i(V_{A}(\theta)+V_{J}(\theta))
\end{bmatrix}\bar{\chi}=0.
\eeqn
Solving the above equation, we can obtain the zero-mode solution
\beqn
\bar{\Psi}(\theta)=\frac{1}{\sqrt{2}}\exp(-\frac{R}{v}\int_{\theta_{c}}^{\theta}V_{A}(\theta')+V_{J}(\theta')d\theta')\begin{bmatrix}
    1 \\ -i
\end{bmatrix}.
\eeqn

\section{C. Detail of vector potential distribution}\label{VP_dis}
In this section, we mainly determine the vector potential distribution under a magnetic field along the y-direction. The supercurrent can be described by the London equation
\beqn
\vec{j}=-\frac{n_{S}e^{2}}{m^{*}}\vec{A}
\eeqn
in which $n_{S}$ is the density of the Cooper pairs in the superconductor. Working into London gauge $\bm{\nabla} \cdot \vec{A} = 0$~\cite{London1935, Bardeen1951}, the vector potential take the form as
\beqn
\vec{A}=A_{y}(z)\vec{y}
\eeqn
satisfying Maxwell equations
\beqn
\nabla^{2}\vec{A}=\frac{1}{\lambda_{L}^{2}}\vec{A}.
\eeqn
with $\lambda_{L}$ is the London penetration length. Thus, 
the general solution can be expressed as
\beqn
A_{y}=C_{1}e^{\frac{z}{\lambda_{L}}}+C_{2}e^{-\frac{z}{\lambda_{L}}}.
\eeqn
According to the boundary conditions  
\beqn
-\partial_{z}A_{y}(z=d/2)=B;~-\partial_{z}A_{x}(z=d/2+h)=B
\eeqn
in the top superconductor  layer ($d/2<z<d/2+h$) and 
\beqn
-\partial_{z}A_{y}(z=-d/2)=B;~-\partial_{z}A_{x}(z=-d/2-h)=B
\eeqn
in the bottom superconductor layer $-d/2-h<z<-d/2$ , the vector potential can be expressed as
\beqn
A_{y}=~-\frac{B\lambda_{L}\sinh{(\frac{z-d/2}{\lambda_{L}}-\frac{h}{2\lambda_{L}})}}{\cosh{\frac{h}{2\lambda_{L}}}}
\eeqn
in the top superconductor and 
\beqn
A_{y}=-\frac{B\lambda_{L}\sinh{(\frac{z+d/2}{\lambda_{L}}+\frac{h}{2\lambda_{L}})}}{\cosh{\frac{h}{2\lambda_{L}}}}
\eeqn
in the bottom superconductor. At the interface between SC and TI $z=\pm d/2$, the vector potential take the value as $A_{y}=\pm B\lambda_{L}\tanh{\frac{h}{2\lambda_{L}}}\approx\pm B\lambda_{L}$ when $h>>\lambda_{L}$. Consequently, according to the continuous distribution, the vector potential in TI region satisfy the boundary condition
\beqn
&&A_{y}(z=\pm d/2) = \pm B\lambda_{L};\nn
&&A_{z}(z=\pm d/2) = 0.
\eeqn 
Thus, the ansatz of the solution into TI region can be expressed as
\beqn
A_{y}&=&-f(y)g'(z)+2B\lambda_{L}\frac{z}{d};\nn
A_{z}&=&f'(y)g(z).
\eeqn
with $f(y)$ and $g(z)$ are the position dependent function, and $g(z=\pm L_{z})=g'(z=\pm L_{z})=0$, $f(y)$ is even function. According to the integration along the closed loop as schematic plot in Fig.1 in the main text, the flux into the region can satisfy
\beqn
B2dR\sin\theta=\oint\vec{A}d\vec{l}=2(-B\lambda_{L}2R\sin\theta+\int_{-d/2}^{d/2}A_{z}(y=R\sin\theta) dz).
\eeqn
Solving above equation, it is easy to obtain that $\delta\Phi(\theta)=\int_{-d/2}^{d/2}A_{z}(y=R\sin\theta) dz=B(2\lambda_{L}+d)R\sin\theta$. Thus, the hopping from the bottom to the top layer gains an additional $\theta$-dependent phase
\beqn
t(r)c_{t}^{\dag}c_{b}\rightarrow t(r)e^{-\frac{e}{\hbar}\int_{-d/2}^{d/2}A_{z}(y=R\sin\theta) dz}c_{t}^{\dag}c_{b}=t(r)e^{-\pi\delta\Phi(\theta)/\Phi_{0}}=t(r)e^{-\pi\eta\sin\theta}
\eeqn
with $\Phi_{0}=h/2e$ ($e>0$) the magnetic quantum flux and $\eta=\pi BR(2\lambda_{L}+d)/\Phi_{0}$ a unitless parameter. This indicates the additional tunneling phase independent of the distribution of flux.

\begin{figure}[htbp]
    \centering
\includegraphics[width=0.9\columnwidth]{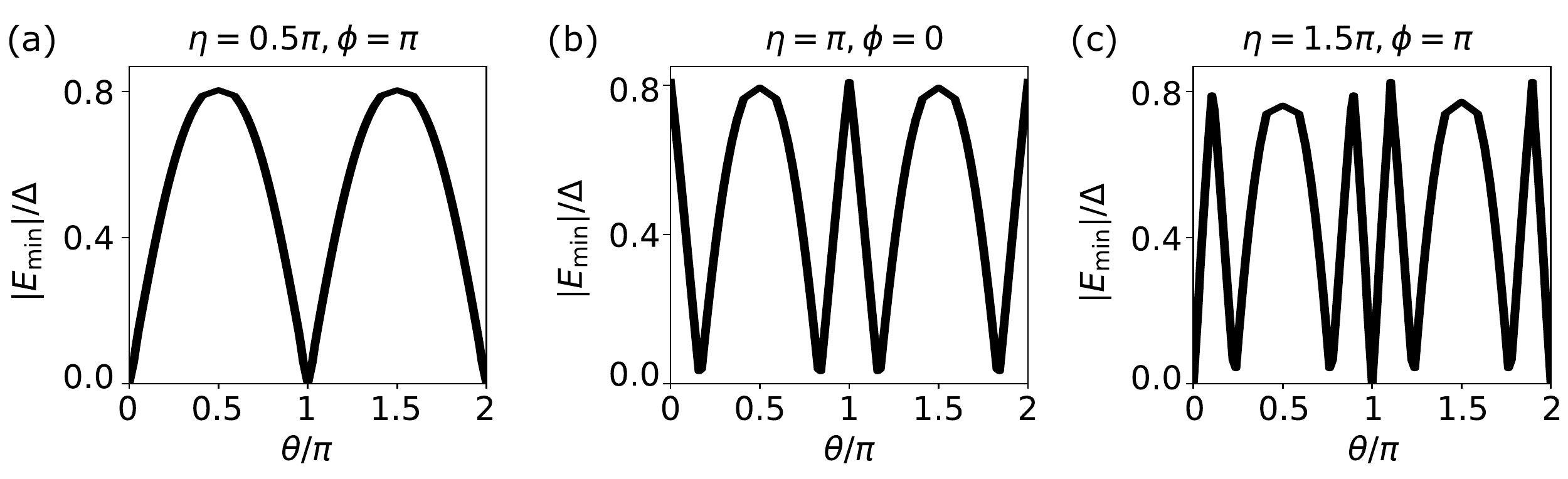}
    \caption{Given d and R: (a-c) are the minimum absolute eigenvalue for different $\eta$ and $\phi$.}
    \label{SFigure_2}
\end{figure}

\section{D. Results independent on specific $f(y), g(z)$}
In this section, we consider another form of $f(y)$ and $g(z)$ to numerically confirm our analysis in the text. We remove the two SC layers and add superconducting pairing into the TI region. In this case, without a magnetic field, the model Hamiltonian of the junction takes the form as
\beqn
H=\begin{bmatrix}
    H_{\rm TI}  & \Delta\big[\Theta(z-w/2)+\Theta(-z-w/2)e^{-i\phi}\big]\\
    \Delta\big[\Theta(z-w/2)+\Theta(-z-w/2)e^{i\phi}\big) & -H_{\rm TI}
\end{bmatrix}
\eeqn
with $i=x,y,z$, the Pauli matrices $\tau, \sigma, s$ acting on the Nambu, orbital and spin space, $M$, $t_{i}$, $\alpha_{i}$ the bulk gap, kinetic energy, and the spin-orbital coupling strength, $\Delta$ and $\phi$ the magnitude and phase of the SC, $w$ the length of the normal region, $\Theta$ Heaviside step function, the superconducting pairing term. Applying 
 the magnetic field along the x-direction, we consider the positive dependent function $f(y)$ and $g(z)$ as
\beqn
f(y)=\bar{B}\sech{\frac{y}{y_{0}}} ; \ \  \text{and} \ \
g(z)=\sech^{2}(\frac{z}{z_{0}})
\eeqn
with 
\beqn
\bar{B}=BR\sin\theta\frac{2\lambda_{L}+d}{-2z_{0}/y_{0}\sech(R\sin\theta/y_{0})\tanh(R\sin\theta/y_{0})\tanh(d/2z_{0})}
\eeqn
$R/y_{0}\rightarrow0, d/z_{0}>>1$. As discussed in the main text, Majorana zero modes can be generated at the mass sign change. We plot the absolute value of the minimal eigen-energy for $\eta=0.5\pi, \phi=\pi$, $\eta=\pi, \phi=0$ and $\eta=1.5\pi, \phi=\pi$ in Fig.~\ref{SFigure_2}. Fig.~\ref{SFigure_2}(a-c) indicate the mass sign change two, four, and six times which are consistent with the analysis in the main text. That is to say, our results are independent of the specific form of $f(y)$ and $g(z)$.

\end{widetext}
\end{appendix}

\end{document}